\documentclass[twocolumn,showpacs,showkeys,superscriptaddress]{revtex4}
\usepackage{graphicx}
\usepackage{bm}
\usepackage{color}
\usepackage{amsmath}
\usepackage{amsfonts}
\usepackage{amssymb}
\usepackage{natbib}
\usepackage{latexsym}

\begin{document}

\title{Magnetic Connections in Curved Spacetime}
\author{Felipe A. Asenjo}
\email{felipe.asenjo@uai.cl}
\affiliation{Facultad de Ingenier\'{\i}a y Ciencias, Universidad Adolfo Ib\'a\~nez, Santiago 7941169, Chile.}
\author{Luca Comisso}
\email{lcomisso@princeton.edu}
\affiliation{Department of Astrophysical Sciences and Princeton Plasma Physics Laboratory, Princeton University, Princeton, NJ 08544, USA.}


\begin{abstract}
The ideal MHD theorem on the conservation of the magnetic connections between plasma elements is generalized to relativistic plasmas in curved spacetime. The connections between plasma elements, which are established by a covariant connection equation, display a particularly complex structure in curved spacetime. Nevertheless, it is shown that these connections can be interpreted in terms of magnetic field lines alone by adopting a $3+1$ foliation of spacetime.
\end{abstract}


\maketitle

\section{Introduction}

A fundamental property of ideal magnetohydrodynamic (MHD) plasmas is that two plasma elements connected by a magnetic field line at a given time will remain connected by a magnetic field line at any subsequent time. This idea, which was analysed in detail by Newcomb \cite{Newcomb} for non-relativistic and special relativistic plasmas, has been the source of great insight into the behavior of such plasmas. 

The importance of the magnetic field line connectivity stem from the fact that it imposes strong constraints on the plasma dynamics, in addition to providing the basis for concepts such as magnetic field line motion \cite{Newcomb} and magnetic topology \cite{Hornig97}. The process of the magnetic reconnection itself, which is thought to power some of the most energetic astrophysical phenomena in the Universe by allowing rapid magnetic energy conversion rates (e.g., Refs. \cite{TajimaShibata1997,Daugh_2009,Sironi2014,Comisso16}), relies on the local violation of these magnetic connections (due to non-ideal effects such as plasma resistivity). Therefore, it is clear that the understanding of the magnetic field line connectivity has significant implications in a variety of astrophysical systems. 

Fostered by recent extensive investigations on the dynamics of relativistic plasmas, the special relativistic formulation of the connection concept was reconsidered by Pegoraro \cite{Pego}, who showed how to cast this ideal MHD property in terms of magnetic field lines alone by means of a time-resetting procedure. Furthermore, it was shown in Refs. \cite{fazcom1,fazcom2} that more general field line connections can persist even in non-ideal relativistic plasmas, setting important constraints on the plasma dynamics by forbidding transitions between configurations with different connectivity.

However, one might wonder if the preservation of the magnetic field line connectivity remains valid in the presence of significant gravitational fields. Examples of plasmas where general relativistic effects are important are those around black holes \cite{ruffini,krolik,Koide2002,Koide2010,Zam,asenjoGRLuca} or in early Universe \cite{TajimaShibata1997,holcomb,Subra,Son}. 
In such cases, general relativity must be taken into consideration in the plasma dynamics. Therefore, the purpose of this work is to investigate whether the magnetic connection concept can be applied to plasmas in curved spacetimes. This is an important inquiry, as in any astrophysical plasma where the gravitational fields are relevant to the dynamics, the fundamental notion of magnetic connection must be valid in order to properly define magnetic reconnection.

In the rest of this manuscript, we will show that the ideal MHD theorem on the preservation of the magnetic connections can be extended to plasmas in a general curved spacetime. This proof allows us to set the basis for the definitions of magnetic topology and magnetic reconnection in high-energy plasmas where general relativity is important.

\section{Connection equation in curved spacetime}

The ideal Ohm's law for plasmas in curved spacetimes is given by the relation
\begin{equation}\label{Ohm1}
U^\nu F_{\mu\nu}=0\, .
\end{equation}
While this equation looks like its counterpart in flat spacetime,  we should emphasize that the spacetime curvature plays an essential role in the proper nature of the electromagnetic and fluid four-velocity fields. The four-velocity $U^\nu$ contains information about the curvature of spacetime through its normalization
\begin{equation}
U^\mu U_\mu=g_{\mu\nu} U^\mu U^\nu=-1\, ,
\end{equation}
where $g_{\mu\nu}$ is the metric tensor with signature $(-,+,+,+)$. Similarly, 
the electromagnetic field tensor  $F^{\mu\nu}=\nabla^\mu A^\nu-\nabla^\nu A^\mu$ is now given in terms of contravariant derivatives $\nabla^\mu=g^{\mu\nu}\nabla_\nu$ and covariant derivatives $\nabla_\mu$. In general, covariant derivatives do not commute \cite{misner}, and thus it is convenient to deal with the covariant version of the electromagnetic field tensor
\begin{equation}\label{EMfield}
F_{\mu\nu}=\nabla_\mu A_\nu-\nabla_\nu A_\mu=\partial_\mu A_\nu-\partial_\nu A_\mu\, .
\end{equation}

An important implication of the ideal Ohm's law \eqref{Ohm1} is that the electromagnetic field is Lie-dragged by the fluid motion. This can be shown in the following way. We substitute Eq. \eqref{EMfield} into the ideal Ohm's law \eqref{Ohm1}, which yields
\begin{equation} \label{Ohm2}
\frac{D{A_\mu} }{D\tau} = U^\nu \nabla_\mu A_\nu \, ,
\end{equation}
where $D/D\tau\equiv U^\nu \nabla_\nu$ represents the convective derivative along the fluid motion. 
Then, taking advantage of the above relations, we write the convective derivative in curved spacetime of $F_{\mu \nu}$ as
\begin{eqnarray}\label{convF1}
{\frac{D}{D\tau}} F_{\mu \nu}&=& U^\alpha \nabla_\alpha (\nabla_\mu A_\nu - \nabla_\nu A_\mu) \nonumber\\
&=& U^\alpha {\nabla _\mu }{\nabla _\alpha }{A_\nu } - U^\alpha  {\nabla _\nu }{\nabla _\alpha }{A_\mu } \nonumber\\
&&+ U^\alpha R_{\beta \nu \mu \alpha} A^\beta - U^\alpha R_{\beta \mu \nu \alpha} A^\beta  \, ,
\end{eqnarray}
where the second line takes into account the non commutative properties of the covariant derivatives.
Using the  Bianchi identity for the Riemann curvature tensor, $R_{\beta \nu \mu \alpha}  +  R_{\beta \mu \alpha \nu} + R_{\beta \alpha \nu \mu} = 0$,
we can rewrite Eq.~\eqref{convF1} as
\begin{eqnarray} \label{convF2}
{\frac{D}{D\tau}} F_{\mu \nu}&=& {\nabla _\mu }({U^\alpha }{\nabla _\alpha }{A_\nu }) - {\nabla _\mu }{U^\alpha }{\nabla _\alpha } A_\nu \nonumber\\
&&- {\nabla _\nu }(U^\alpha {\nabla_\alpha} A_\mu) + {\nabla_\nu} U^\alpha {\nabla_\alpha} A_\mu \nonumber\\
&&- U^\alpha  R_{\beta \alpha \nu \mu}  A^\beta  \, .
\end{eqnarray}
Substituting $\nabla_\alpha A_\nu = F_{\alpha \nu}  + \nabla_\nu A_\alpha$ in the above equation, and making use of Eq. \eqref{Ohm2}, we obtain
\begin{eqnarray}\label{convF3}
{\frac{D}{D\tau}} F_{\mu \nu}&=&   {\nabla_\nu} U^\alpha F_{\alpha \mu}  - \nabla_\mu U^\alpha F_{\alpha \nu}+U^\alpha \nabla_\mu \nabla_\nu A_\alpha \nonumber\\
&& - U^\alpha \nabla_\nu \nabla_\mu A_\alpha  - U^\alpha  R_{\beta \alpha \nu \mu}  A^\beta  \, .
\end{eqnarray}
Finally, exploiting the non-commutative properties of the convective derivatives, it is straightforward to find that
\begin{equation} \label{eq_F}
{\frac{D}{D\tau}} F_{\mu \nu}  = \nabla_\mu U^\alpha F_{\nu \alpha}   -  \nabla_\nu U^\alpha F_{\mu \alpha}\, ,
\end{equation}
implying that the electromagnetic field is Lie-dragged with the fluid in the ideal MHD description. A different derivation, which led to the same conclusion, was given by Achterberg \cite{acht83}.

Here, we intend to take a step further by showing how two different fluid elements are connected to each other if the ideal Ohm's law is satisfied. For this purpose, we introduce the four-displacement $\Delta x^\mu$ of a given fluid element, which is related to the fluid four-velocity as
\begin{equation}
{\Delta x^\mu} = U^\mu {\Delta \tau} \, ,
\end{equation}
where $\Delta \tau$ is the variation of the proper time. Thereby, two different fluid elements are separated by a spacelike
event-separation four-vector $d l^\mu = x'^{\mu} - x^{\mu}$, which establishes simultaneity between events when $dl^0=0$ \cite{Pego}. This event-separation four-vector is transported by the fluid motion, and its convective derivative in curved spacetime can be calculated to be
\begin{eqnarray} \label{convsep1}
{\frac{D}{D\tau}} d l^\mu  &=& U^\alpha \nabla_\alpha x'^{\mu} -  U^\alpha \nabla_\alpha x^{\mu}   \nonumber\\
&=& U^\alpha \partial_\alpha x'^{\mu} - U^\alpha \partial_\alpha x^{\mu} + U^\alpha {\Gamma^\mu}_{\alpha \lambda} x'^{\lambda}  -  U^\alpha {\Gamma^\mu}_{\alpha \lambda} x^\lambda   \nonumber\\
&=&  U'^{\mu} - U^{\mu} + U^\alpha {\Gamma^\mu}_{\alpha \lambda} d{l^\lambda}  \, ,
\end{eqnarray}
where ${\Gamma^\mu}_{\alpha \lambda}$ are the Christoffel symbols associated to the metric $g_{\mu\nu}$. Recalling that $x'^{\alpha} = x^{\alpha} + dl^{\alpha}$, we have $U'^{\mu}=U^{\mu}(x_\phi + d l_\phi)$. Moreover, from the definition of derivative, $U^{\mu}(x_\alpha + d l_\alpha) - U^{\mu}(x_\alpha) = d l_\alpha {\partial {U^\mu}(x_\alpha)}/{\partial {x^\alpha}}$ as $d l_\phi \to 0$. Therefore, we can rewrite Eq.~\eqref{convsep1} as
\begin{equation} \label{eq_dl}
{\frac{D}{D\tau}} d l^\mu  = d {l^\lambda} \nabla_\lambda U^{\mu}  \, ,
\end{equation}
which shows how the event-separation four-vector dynamically propagates along the fluid motion.

We are now in the position to show that if the event-separation four-vector $d{l^\mu}$ is chosen in such a way that $d{l^\mu} F_{\mu \nu}$ is initially zero, then $d{l^\mu} F_{\mu \nu}$ is always zero. Indeed, by writing the convective derivative in curved spacetime of the quantity $d l^\lambda F_{\mu \nu}$ as
\begin{equation} \label{}
{\frac{D}{D\tau}} \left( d l^\mu F_{\mu \nu} \right)  =  U^\alpha \nabla_\alpha d{l^\mu} F_{\mu \nu} +  d{l^\mu} U^\alpha \nabla_\alpha F_{\mu \nu} \, ,
\end{equation}
we can directly use Eqs.~\eqref{eq_F} and \eqref{eq_dl} to obtain 
\begin{eqnarray} 
{\frac{D}{D\tau}}\left( d l^\mu F_{\mu \nu} \right) &=&  d{l^\alpha} \nabla _\alpha U^\mu  F_{\mu \nu} \nonumber\\
&&+  d{l^\mu}\left( \nabla_\mu U^\alpha F_{\nu \alpha}    -  \nabla_\nu U^\alpha F_{\mu\alpha} \right) \, ,
\end{eqnarray}
which leads us to the {\it connection equation} 
\begin{equation} \label{connection_eq}
{\frac{D}{D\tau}} \left( d l^\mu F_{\mu \nu} \right)  =  -  \left(\nabla_\nu U^\alpha\right) \left(d{l^\mu} F_{\mu \alpha}\right) \, .
\end{equation}
From this equation it follows that if initially we have 
\begin{equation}\label{condFroz}
d{l^\mu} F_{\mu \alpha} = 0 \, ,
\end{equation}
then $D(d l^\mu F_{\mu \nu})/{D\tau} = 0$ at every time, implying that $d{l^\mu} F_{\mu \alpha}$ will remain null at all times (regularity properties of the four-velocity field $U^\alpha$ are assumed). This mathematical statement represents a generalization of the ideal MHD theorem on the conservation of the magnetic connections between plasma elements \cite{Newcomb} for a relativistic plasma in curved spacetime.

Eq. \eqref{connection_eq} generalizes the connection equation in flat spacetime derived in Ref. \cite{Pego}. It provides also a path for generalizing the special relativistic extended MHD theory developed in Refs.~\cite{fazcom1,fazcom2} to curved spacetime. However, while the connection equation \eqref{connection_eq} in the flat spacetime limit has a well-defined meaning, its interpretation in curved spacetime is more subtle. In flat spacetime,  if there is simultaneity between two events, the condition \eqref{condFroz} yields the vectorial conditions $d\vec l\cdot \vec E=0$ and $d\vec l\times \vec B=0$, the latter of which implies that the connection between plasma elements is established by the magnetic field lines. This result is generally expressed by saying that the magnetic field lines are ``frozen" into the plasma. On the other hand, the condition \eqref{condFroz} has more complex implications in curved spacetime, and its interpretation in terms of magnetic field lines alone requires a specific choice in the way that the electric and magnetic four-vector fields are defined in an arbitrary spacetime.

\section{Magnetic connections}

In order to formulate this generalization of the ideal MHD frozen-in theorem \cite{Newcomb} in terms of magnetic field connections, we analize Eqs.~\eqref{connection_eq} and \eqref{condFroz} using two different definitions of the electric and magnetic four-vectors. By taking projections of the electromagnetic field tensor onto different hypersurfaces, we highlight the conditions that allow us to specify the connection concept in curved spacetime in terms of magnetic field lines alone.

\subsection{Electromagnetic field projection onto the fluid}

It is often convenient to define covariant electric and magnetic four-vectors by projecting the electromagnetic field onto  hypersurfaces orthogonal to the fluid four-velocity. In this way, the aforementioned four-vectors reduce to the electric and magnetic fields in a comoving plasma frame. Following this approach (e.g. Refs.~\cite{lich,beke1,anile,acht83,beke2,kawa14,davignon,pegoJ}), the electric and magnetic four-vectors can be defined as
\begin{equation} \label{defiBeke}
\mathcal{E}_\mu = F_{\mu \nu} U^\nu  \, ,  \qquad \mathcal{B}_\mu = \frac{1}{2} \epsilon_{\mu \nu \alpha \beta} U^\nu F^{\alpha \beta}\, ,
\end{equation}
where  $\epsilon_{\mu \nu \alpha \beta}$ is the completely antisymmetric Levi-Civita tensor, normalized such that $\epsilon_{0 1 2 3} = \sqrt{-g}$.
From the previous equations, it follows that $\mathcal{B}_\alpha U^\alpha = 0$ and $\mathcal{E}_\alpha U^\alpha = 0$. Therefore, $\mathcal{B}_\alpha$ and $\mathcal{E}_\alpha$ have only three independent components each.
It is important to remark that the time component of $\mathcal{B}_\mu$ is $\mathcal{B}_0 = \tfrac{1}{2} \epsilon_{0ijk} U^i F^{jk}\neq 0$ in general, and thus we have a well-defined magnetic field only when the plasma is at rest with $U^i=0$ (latin indices running from 1 to 3 for spatial components). We will see how this definition has repercussions on the magnetic connection concept.

Since the ideal Ohm's law \eqref{Ohm1} defines a null electric field measured by the comoving observer, $\mathcal{E}_\mu = F_{\mu \nu} U^\nu=0$, the electromagnetic field constructed from the definitions \eqref{defiBeke} can be written just as
\begin{equation} \label{lFbbeke}
F_{\mu \nu} =  \epsilon_{\mu \nu \alpha \beta} U^\alpha \mathcal{B}^\beta  \, .
\end{equation}
Similarly, the dual of the electromagnetic field becomes
\begin{equation} \label{dualFbbeke}
F^*_{\mu \nu} = - U_\mu \mathcal{B}_{\nu} + U_\nu \mathcal{B}_{\mu} \, .
\end{equation}
Note that in this formalism, the evolution of the magnetic four-vector, which can be deduced from the homogeneous Maxwell equation ${\nabla_\mu} F^{* \mu \nu} = 0$, is governed by the equation
\begin{equation} \label{Eq_Evolution_B}
\frac{D {\mathcal{B}^{\nu}}}{D\tau} = ({\nabla_\mu} U^\nu) \mathcal{B}^{\mu} - ({\nabla_\mu} U^\mu) \mathcal{B}^{\nu} +  U^\nu ({\nabla_\mu} \mathcal{B}^{\mu}) \, .
\end{equation}  
This equation differs from the standard magnetic field evolution equation in flat spacetime, since the last term does not vanish as the magnetic four-vector \eqref{defiBeke} is not divergence free in general.

We are now in the position to determine the connection condition \eqref{condFroz} under this formalism. Indeed, using Eq. \eqref{lFbbeke} we duly obtain
\begin{equation}\label{condBekens}
\epsilon_{\mu\nu\alpha\beta}d{l^\nu}U^\alpha g^{\beta\lambda}\mathcal{B}_\lambda = 0\, .
\end{equation}
The above condition must be written in terms of the covariant magnetic four-vector \eqref{defiBeke}, as this is the definition that coincide with the proper concept of magnetic field for an observer at rest.

Let us analyse the condition \eqref{condBekens} in the frame where $dl^0 = 0$. This assumption does not affect the generality of the analysis presented here, since if $dl^0 \neq 0$, one can always restore the simultaneity between the two connected plasma elements by performing the transformation $dl^\mu \rightarrow dl'^\mu = dl^\mu + U^\mu d\lambda$ such that in this reference frame $dl'^\mu = 0$. Indeed, due to the validity of Ohm's law $U^\nu F_{\mu\nu}=0$, this transformation keeps unmodified the connection equation \eqref{connection_eq}. Therefore, we can proceed by evaluating the time component of Eq.~\eqref{condBekens}, which gives a scalar equation with the form
\begin{equation}\label{condBekensTime}
\epsilon_{0ijk}d{l^i}U^j \left(g^{k0}\mathcal{B}_0+g^{km}\mathcal{B}_m\right) = 0\, ,
\end{equation}
where we have explicitly written every component of the metric. In general, a metric can have $g^{k0}\neq 0$, as is the case for the spacetime of a rotating black hole. Also, a general spacetime can have $g^{ij}\neq 0$ for $i\neq j$.
Likewise, the spatial components of Eq.~\eqref{condBekens} produce the equation
\begin{equation}\label{condBekensSpace}
\epsilon_{0ijk}d{l^j}\chi^{km}\mathcal{B}_m = 0\, ,
\end{equation}
where the matrix $\chi$ is defined as
\begin{equation}
\chi^{ij}=g^{ij}+g^{00}\frac{U^iU^j}{(U^0)^2}-\frac{1}{U^0}\left(U^i g^{0j}+U^j g^{0i}\right)\, ,
\end{equation}
and $U^0$ can be obtained from the normalization of the velocity 
\begin{equation}
U^0=-\frac{g_{0i}}{g_{00}}U^i+\sqrt{\left(\frac{g_{0i}}{g_{00}}U^i\right)^2-\frac{g_{ij}}{g_{00}}U^iU^j-\frac{1}{g_{00}}}\, .
\end{equation}

We can further analyse the condition \eqref{condBekensSpace}. While this condition reflects the connectivity statement for a plasma that obeys the ideal Ohm's law \eqref{Ohm1}, its interpretation is not straightforward and in general cannot be done in terms of the standard magnetic field alone. Indeed, Eq. \eqref{condBekensSpace} shows that the connection between plasma elements occurs through the vector field $\chi^{ij}\mathcal{B}_j$ and not the magnetic field. The behavior of the field $\chi^{ij}\mathcal{B}_j$ is rather complicated, and in general depends on the plasma fluid velocity. Furthermore, it is difficult to understand its large dependence on different possible spacetimes where the plasma is moving. 

However, two interesting limits should be noticed from the above equations. The first one is the limit of plasma at rest, i.e. $U^i=0$. In this case, Eq.~\eqref{condBekensTime} is trivially satisfied, while Eq.~\eqref{condBekensSpace} establishes that plasma elements are connected through the field $g^{ij}\mathcal{B}_j$. This result is analogue to the magnetic connections in flat spacetime. 
The second interesting limit is for a symmetric and static spacetime, where $g_{ij}=g_{ii}\delta_{ij}$ and $g_{0i}=0$. Spacetimes that satisfy those conditions are, for example, the Schwarzschild black hole metric, the cosmological Friedman-Lema\^{i}tre-Robertson-Walker metric, wormholes, etc. \cite{misner}. In these cases, Eq. \eqref{condBekensTime} simplifies to $\epsilon_{0ijk}d{l^i}U^jg^{kk}\mathcal{B}_k = 0$, whereas Eq.~\eqref{condBekensSpace} is now written in terms of $\chi^{ij}=g^{ii}\delta^{ij}+g^{00}{U^iU^j}/{(U^0)^2}$. Even in this simplified limit the connections are established by a complex field that depends on the fluid velocity and the spacetime metric.

\subsection{3+1 foliation of spacetime}

In order to avoid the above difficulties, we explore another way to define the electric and magnetic four-vectors. Indeed, as shown by Thorne and Macdonald \cite{TM_82}, there are situations in which it is convenient to implement a $3+1$ decomposition of plasma and electromagnetic fields by projecting every physical vector and tensor onto timelike and spacelike hypersurfaces of the metric, in such a way to obtain a set of plasma equations that resemble those found in special relativity (e.g. Refs.~\cite{TM_82,Thorne86,Zhang89,311,312,314,315,316,317,318,319,320,asenjo1,asenjo2,das1,das2,asenjoGRLuca}). 

To this purpose, let us consider a general spacetime described by the metric \cite{ADM62,misner}
\begin{equation}
ds^2=g_{\mu\nu}dx^\mu dx^\nu=-\alpha^2 dt^2+\gamma_{ij}\left(dx^i+\beta^i dt\right)\left(dx^j+\beta^j dt\right)\, ,
\end{equation}
where $\alpha$ is known as the lapse function, $\beta^ \mu=(0,\beta^i)$ is the shift vector (related to non-static spacetimes), and $\gamma_{ij}$ is the induced three-metric on the spacelike hypersurfaces $\Sigma_t$ of constant time $t$. The timelike unit vector field normal to $\Sigma_t$ is defined by a timelike vector field $n^\mu$ that satisfies the normalization condition $n_\mu n^\mu=-1$. This vector field has the form $n_\mu = - \alpha {\nabla_\mu}t = (-\alpha,0,0,0)$, $n^\mu=(1/\alpha, -\beta^i/\alpha)$ \cite{TM_82,Thorne86}, and can be interpreted as the four-velocity of the local fiducial observer (FIDO), which is at rest in the absolute space. 
The projection tensor, whose spatial components coincide with the components of the three-metric $\gamma_{ij}$, is defined as \cite{TM_82,Thorne86}
\begin{equation} \label{}
{\gamma _{\mu \nu }} = {g_{\mu \nu }} + {n_\mu }{n_\nu } \, .
\end{equation}
Note that the definitions of $n_\mu$ and $\gamma_{\mu \nu}$ satisfy the conditions ${\beta ^\mu }{n_\mu } = 0$ and $n^\mu\gamma_{\mu\nu}=0$. Then, the $3+1$ decomposition is achieved by projecting every vector/tensor onto $n^\mu$ (timelike hypersurfaces) and onto $\gamma _{\mu \nu}$ (spacelike hypersurfaces).

Under this $3+1$ formalism, the electric and magnetic four-vectors are defined as \cite{TM_82,Thorne86}
\begin{equation} \label{emFieldn}
E^\mu = {n_\nu }{F^{\mu \nu }} \, , \qquad
B^\mu = \frac{1}{2}{n_\rho }{\epsilon ^{\rho \mu \sigma \tau }}{F_{\sigma \tau }}  \, .
\end{equation}
In this description, both fields are spacelike vectors since ${n_\mu }{E^\mu } = 0$ and ${n_\mu }{B^\mu } = 0$. Furthermore, we have always $B^0=0$ by the definition of $n_\mu$. Therefore, the magnetic four-vector has a well-behaved magnetic field. On the other hand, in this $3+1$ formalism, the electric field does not vanish. Thus, the three-dimensional expressions of the electric and magnetic fields turn out to be the standard ones (as their counterparts in flat spacetimes). Note that similar decompositions can be performed for other plasma quantities as well. For example, the four-vector plasma fluid velocity can be written as (e.g. Refs.~\cite{asenjo1,asenjo2,das1,das2})
\begin{equation}\label{plasmafluidecomposed}
U^\mu=\alpha\Gamma n^\mu+\Gamma{\gamma^\mu}_\nu v^\nu\, ,
\end{equation}
with the Lorentz factor $\Gamma=[\alpha^2-\gamma_{ij}(\beta^i\beta^j+2\beta^i v^j+v^iv^j)]^{-1/2}$ and $v^\mu=(0,v^i)$, where $v^i$ are the spatial components of the fluid velocity.

From the previous definitions, we can decompose the electromagnetic field tensor as
\begin{equation} \label{f3mas1}
{F^{\mu \nu }} = {E^\mu }{n^\nu } - {E^\nu }{n^\mu } - {\epsilon ^{\mu \nu \rho \sigma }}{B_\rho }{n_\sigma } \, ,
\end{equation}
while its dual can be expressed as
\begin{equation} \label{dualF3mas1}
{F^{\mu \nu }}^* = {B^\mu }{n^\nu } - {B^\nu }{n^\mu } + {\epsilon ^{\mu \nu \rho \sigma }}{E_\rho }{n_\sigma }  \, .
\end{equation}
Using Eq. \eqref{dualF3mas1} in the homogenous Maxwell equation, it is straightforward to obtain a covariant magnetic four-vector field equation that yields the standard 3D version of the magnetic field evolution equation (see, e.g., Ref. \cite{asenjo1}), since in this case $B^\mu$ is divergence free. This indicates that the magnetic four-vector given in Eq. \eqref{emFieldn} is suitable to recast the theorem described in Sec. II in terms of the standard magnetic connection concept.

We can now analyse the connection condition by substituting Eq. \eqref{f3mas1} into Eq. \eqref{condFroz}. This gives us
\begin{equation}\label{condF3mas1}
n_\mu \left(d{l^\nu} E_{\nu}\right)-\epsilon_{\mu\nu\rho\sigma}dl^\nu B^\rho n^\sigma = 0\, ,
\end{equation}
where we have used that $dl^\mu n_\mu \equiv 0$ for simultaneous events (the simultaneity between two connected plasma elements can again be always  obtained by means of the time-resetting $dl^\mu \rightarrow dl'^\mu = dl^\mu + U^\mu d\lambda = 0$ which leaves the connection equation \eqref{connection_eq} unaltered). Therefore, we can analyse the projections of Eq.~\eqref{condF3mas1} onto timelike and spacelike hypersurfaces. In the first case, by contracting Eq.~\eqref{condF3mas1} with $n^\mu$, we find
\begin{equation}\label{condF3mas1Time}
d{l^\nu} E_{\nu}=dl^i E_i= 0\, .
\end{equation}
This implies that the electric field is orthogonal to the event-separation vector, in analogy to ideal MHD plasmas in flat spacetime \cite{Pego}. On the other hand, the spacelike hypersurface projection of Eq.~~\eqref{condF3mas1} yields
\begin{equation}\label{condF3mas1Space}
\epsilon_{0ijk}dl^j B^k = 0\, ,
\end{equation}
as $n^0=1/\alpha$,  $B^0=0$, and $dl^0=0$ for simultaneity. Eq.~\eqref{condF3mas1Space} tells us that, under this $3+1$ foliation of spacetime, the connection between plasma elements occurs through the magnetic field $B^i$. This result reveals that the condition \eqref{condF3mas1} is the natural general relativistic extension of the connection condition for classical and special relativistic plasmas. Therefore, the $3+1$ formalism allows us to interpret the ideal MHD connection theorem in terms of magnetic field lines alone even for relativistic plasmas in curved spacetimes.

Finally, we prove that the condition \eqref{condF3mas1Time} is consistent with the ideal Ohm's law. Substituting the electromagnetic field tensor \eqref{f3mas1} and the fluid four-velocity \eqref{plasmafluidecomposed} into the Ohm's law \eqref{Ohm1}, we get
\begin{equation}
\alpha E_\nu-n_\nu {\gamma^\mu}_\beta v^\beta E_\mu-\epsilon_{\mu\nu\rho\sigma}{\gamma^\mu}_\beta v^\beta B^\rho n^\sigma=0\, .
\end{equation}
Contracting it with $dl^\nu$ we obtain
\begin{equation}
 \alpha\, dl^\nu E_\nu=\epsilon_{\mu\nu\rho\sigma}dl^\nu{\gamma^\mu}_\beta v^\beta B^\rho n^\sigma\, .
\end{equation}
Both sides of the above equation vanish, the left-hand side by Eq.~\eqref{condF3mas1Time}, while  the right-hand side by Eq.~\eqref{condF3mas1Space}. Thus, the above equation is identically satisfied.

\section{Discussion}

The connection equation \eqref{connection_eq}, with its solution \eqref{condFroz}, allow us to prove that, in a general curved spacetime, there exist connections between plasma elements that are dynamically preserved if the plasma satisfies the ideal Ohm's law.

The identification of these connections with properly defined magnetic connections is not straightforward and must be worked out carefully.
In flat spacetime, the magnetic connection concept is well-defined when we refer to a frame where the connected elements are simultaneous. This continue to be the case even when the reference frame is changed, since the validity of the ideal Ohm's law allow us to reset the time in such a way to restore simultaneity without changing the connectivity of the plasma elements \cite{Pego}. Therefore, fundamental concepts introduced for non-relativistic plasmas, such as magnetic field line motion, magnetic topology, and magnetic reconnection, can be adopted also in special relativistic regimes. 

In order to apply these concepts to general relativistic plasmas, the magnetic four-vector field has to be defined in such a way to recover the standard notion of magnetic connection. We have shown that this can be done through a $3+1$ foliation of spacetime. Indeed, the magnetic four-vector defined in \eqref{emFieldn} allows us to maintain, in a curved spacetime analogue fashion, the concepts related to the magnetic field line connectivity that have been adopted for non-relativistic and special relativistic plasmas.
On the contrary, if other definitions of the magnetic four-vector field are invoked, the standard magnetic connection concept is not guaranteed to hold. This is the case for the definitions \eqref{defiBeke}, where the electric and magnetic fields measured in a comoving plasma frame are considered. These different definitions lead to a redefinition of the connected fields that do not coincide in general with the magnetic connections (a part for very specific cases as the limit of plasma at rest).

On account of these reasons, the $3+1$ decomposition of the electromagnetic and plasma quantities results to be the most suitable approach to formulate an ideal MHD frozen-in theorem in general relativity. This formalism also provides a straightforward way to generalize the magnetic connection hypersurfaces \cite{GrallaJacob,pegoJ}, which are 2D hypersurfaces that satisfy the connection condition \eqref{condFroz} and reduce to magnetic connection lines in a chosen reference frame when taking sections of these surfaces at a fixed time. Indeed, substituting the decomposed electromagnetic field \eqref{f3mas1} into the ideal Ohm's law \eqref{Ohm1}, we find
\begin{equation}
U^\mu F_{\mu\nu}=\left(U^\mu n_\mu\right) E_\nu-n_\nu\left(U^\mu E_\mu\right)-\epsilon_{\mu\nu\alpha\beta}U^\mu B^\alpha n^\beta=0\, .
\end{equation}
Contracting this equation with $n^\nu$ first, and then with $B^\nu$, we get
\begin{equation}
U^\mu E_\mu=0\, ,\qquad B^\mu E_\mu=0\, .
\end{equation}
Thus, for a plasma that obeys the ideal Ohm's law, the electric and magnetic four-vectors are orthogonal. The same is true for the electric and fluid velocity four-vectors. These two conditions allow to calculate $B^\mu F_{\mu\nu}$, which turns out to vanish
\begin{equation}
B^\mu F_{\mu\nu}=\left(B^\mu n_\mu\right) E_\nu-n_\nu\left(B^\mu E_\mu\right)-\epsilon_{\mu\nu\alpha\beta}B^\mu B^\alpha n^\beta \equiv 0\, .
\end{equation}
Therefore, $U^\mu$ and $B^\mu$ are orthogonal to the electromagnetic field. Hence, the four-vector event separation $dl^\mu$ lies in the hypersurface formed by the four-vectors $U^\mu$ and $B^\mu$. This implies that the magnetic field lines are organized on magnetic field hypersurfaces which satisfy the connection equation in any reference frame.

We should emphasize that the proof of the conservation of the magnetic connections between plasma elements relies only on the validity of the ideal Ohm's law and the homogeneous Maxwell equation.  Hence, the magnetic field line connectivity is preserved under more general conditions than those required for ideal MHD to hold.

 Finally, we remark that the presented treatment of the magnetic connection concept provides an appropriate framework to study magnetic reconnection in curved spacetime \cite{asenjoGRLuca,LucaGRasenjo}. Accordingly, it is the magnetic field defined by Eq.~\eqref{emFieldn}, and not another quantity, which can be considered to reconnect in general relativistic systems. For that reason, we believe that this analysis can help in deepening our understanding of magnetic reconnection in high-energy astrophysical plasmas.

\begin{acknowledgments}
We acknowledge fruitful discussions with Russell Kulsrud and Yajie Yuan. 
F.A.A. thanks Fondecyt-Chile Grant No. 11140025. 
L.C. is grateful for the hospitality of the Universidad Adolfo Ib\'a\~nez, where part of this work was done. 
\end{acknowledgments}

\end{document}